# On the Spaces and Dimensions of Geographical Systems

Yanguang Chen

(Department of Geography, College of Urban and Environmental Sciences, Peking University, Beijing, 100871 P.R. China)

**Abstract**: To remove the confusion of concepts about different sorts of geographical space and dimension, a new framework of space theory is proposed in this paper. Based on three sets of fractal dimensions, the geographical space is divided into three types: real space (R-space), phase space (P-space), and order space (O-space). The real space is concrete or visual space, the fractal dimension of which can be evaluated through digital maps or remotely sensed images. The phase space and order space are both abstract space, the fractal dimension values of which cannot be estimated with one or two maps or images. The dimension of phase space can be computed by using time series, and that of order space can be determined with cross-sectional data in certain time. Three examples are offered to illustrate the three types of spaces and fractal dimension of geographical systems. The new space theory can be employed to explain the parameters of geographical scaling laws, such as the scaling exponent of the allometric growth law of cities and the fractal dimension based on Horton's laws of rivers.

## 1. Introduction

Fractals provide a new ways of looking at geographical phenomena and a new tool for geographical spatial analysis. Since fractal geometry came out, many of our theories in human and physical geography have been being reinterpreted by using concepts from fractals (Batty, 1992). Though we make great progress in theoretical exploration of geographical fractals, we meet many puzzling questions which cannot be answered in terms of traditional notion of space. For example, if the dimension of urban boundaries is estimated by means of the area-perimeter scaling relation, the value is expected to come between 1 and 2. However, some cases make exceptions, and the results are less than 1 or greater than 2. The fractal dimension of river systems is another pending question (Chen, 2009). Evaluating the fractal dimension of a river network with Horton's laws, we can find that the values sometimes go beyond the lower limit 1 and upper limit 2 (LaBarbera and Rosso, 1989). In short, lots of dimension problems associated with power laws in geography such as the law of allometric growth have been puzzling geographers for a long time.

The root of the problems lies in the concept of geographical space. We confuse the real space which can be recognized directly and the abstract space which cannot be perceived directly. When we calculate the fractal dimension of a city's form through a digital map, we face a real geographical space, which can be felt visually. However, if we study the rank-size distribution of cities, we will meet fractal dimensions which are not subject to real geographical space. Similarly, the fractal dimension of river systems based on Horton's laws is actually defined in an abstract space rather than in real space. Therefore, the fractal dimension values cannot be always confined by the topological dimension and Euclidean dimension of the embedding space in which a fractal body exists (see Vicsek, 1989 for the concept of embedding space).

In a sense, geography is a science on geographical space. Geographical space used to be defined by distance variable (Johnston, 2003). However, because of scaling invariance of geographical phenomena, distance is not always sufficient for us to define a space. It is necessary to employ the concept of dimension to make spatial analyses. Space and dimension are two conjugated concepts. Where there is a concept of space, there is a concept of dimension, and *vice versa*. Fractal dimension is the dimension of space, and geographical space is the space with certain dimension. The notion of dimension used to be in the domain of theoretical science. However, because of fractal geometry, the dimension concept evolves from theoretical sciences into empirical sciences. Many geographical phenomena or even geography itself have fractal nature (Goodchild and Mark, 1987; Batty, 1992). It is essential for us to make geo-spatial analyses by using the ideas of fractal dimension. Now, in both physical geography and human geography, we meet a kind of fractal dimensions which fail to measure the real geographical space. This suggests that there is generalized geographical space which cannot be felt visually and cannot be characterized with the common fractal dimension. Thus, we have to face new problems: what is the abstract space, and how to understand it and measure it?

In this paper, geographical spaces are classified as three types in theoretical perspective, and the fractal dimensions in a broad sense corresponding to these spaces are put in order. Section 2 presents new definitions of geographical spaces and dimensions based on allometric scaling relations, section 3 gives several examples to illustrate these spaces and dimensions, and section 4 shows how to apply these notions to concrete geographical scaling analysis. Finally, the paper is concluded with summary and remarks.

## 2. Definitions

### 2.1. Allometric growth, fractal dimension, and generalized space

The concepts of geographical space and dimension can be demonstrated by means of the law of allometric growth. The problem of dimension coming from the allometric relations that once puzzled many geographers (Lee, 1989). Today, the allometric scaling relation is very useful in geographical analysis, especially in urban spatial analysis (Batty, 2008; Batty *et al*, 2008; Bettencourt *et al*, 2007; Chen, 2010a; Samaniego and Moses, 2008). In fact, the allometric scaling exponent is associated with fractal dimension (Batty and Longley, 1994; Chen, 2010a). Based on the fractal notion defined by Mandelbrot (1983), we can comprehend geographical fractals from two perspectives. One is *scaling invariance* (scale-free), or scaling symmetry, which can be formulated as a functional equation

$$f(\lambda x) \propto \lambda^{\pm \alpha} f(x), \tag{1}$$

where $\alpha$ refers to the scaling exponent, and $\lambda$ to the scale factor. The other is *dimensional consistency/homogeneity*, which can be formulated as geometrical measure relations between line ($L$, $d=1$), area ($A$, $d=2$), volume ($V$, $d=3$), and generalized volume ($M$, $d$) in the form

$$L^{1/1} \propto A^{1/2} \propto V^{1/3} \propto M^{1/d_f}, \tag{2}$$

where $d_f$ denotes an Euclidean or fractal dimension. This implies that a measure will be proportional to another one if and only if the two measures share the same dimension.

Now let's take the allometric relation between urban area ($S$) and population ($P$) as an example to show how to understand the fractal concept in urban geography. This example can be generalized to physical domain. The allometric growth law indicates that the rate of relative growth of urban area is a constant fraction of the rate of relative growth of urban population. The allometric equation can be expressed as

$$S(x) \propto P(x)^b, \qquad (3)$$

where $b$ is a scaling exponent, $x$ represents **t**ime ($x=t$) in the dynamic process of urban evolvement, or denotes ran**k** ($x=k$) in a hierarchy of cities, or refers to **r**adius ($x=r$) of city form. No matter what $x$ is, for a given **t**ime or a specified ran**k**, the measures $S$ and $P$ can be connected with one certain city. Let $b=D_s/D_p$. Introducing a common scale $R$ with dimension $d=1$ into equation (3), we have

$$S(R)^{1/D_s} \propto P(R)^{1/D_p} \propto R. \qquad (4)$$

where $R$ is the maximum average **r**adius of urban area. The geometric measure relation can be decomposed as follows

$$S(R) \propto R^{D_s}, \qquad (5)$$

$$P(R) \propto R^{D_p}. \qquad (6)$$

Considering the scale-free nature of urban form as fractals (Batty and Longley, 1994; Frankhauser, 1994), we cannot determine the value of the maximum radius in theory. Thus we should adopt a variable **r**adius $r$ to replace the constant $R$, then equations (5) and (6) change to the following power laws

$$S(r) \propto r^{D_s}, \qquad (7)$$

$$P(r) \propto r^{D_p}, \qquad (8)$$

where $D_s$ refers to the fractal dimension of urban form, and $D_p$ to the fractal dimension of urban population.

If we reverse the path and go from equations (7) and (8) to equation (3), we can see that the scaling exponent of allometric relations is the ratio of fractal dimension of urban form to that of urban population. The dimension values may be different from city to city, but the ratios tend to be equal on the average. On the other hand, the fractal dimension is a spatial uniformity index: a geographical phenomenon has a higher dimension value the more homogeneous it becomes. Therefore, if urban land use area grows rapider than population does, it will show a result $b>1$ according to equation (3), and thus $D_s>D_p$, which just consists with the given relation $b=D_s/D_p$.

To sum up, if $x=r$, we have a fractal dimension in its narrow sense; however, if $x=t$ or $x=k$, the dimension does not belong to real geographical space commonly understood by geographers, the parameters $b=D_s/D_p$ is in fact the dimensions ratio in an abstract space. In other words, the dimension based on time series or hierarchical structure is different from that based on network or pure spatial form. In order to avoid confusing concepts, it is necessary to define new types of geographical space to adopt these dimensions.

## 2.2. RPO: Definition of generalized geographical space

For explaining the fractal dimension of both the human systems (e.g. urban systems) and physical systems (e.g. river systems), we'd better categorize the geographical spaces into three types: real space (R-space), phase space (P-space), and order space (O-space).

1. Real space. This is the common geographical space which can be reflected by maps or remote sensing images (RSI), and so forth. The real form, relation, location, and neighborhood of geographical phenomena are always given in this kind of spatial way. Real space can be represented with Cartesian coordinates systems familiar to geographers.

2. Phase space. It can be described by time series of geographical systems. Phase space is a concept taken from physics, but it is very helpful in reflecting the regularity in the temporal series of geographical evolution. For a physical system with $n$ degrees of freedom, we can define a $2n$-dimensional space with coordinates $(x_1, x_2,…, x_n, y_1, y_2, …, y_n)$, where $x_i$ describes the degrees of freedom of the system and $y_i$ are the corresponding momenta ($i=1,2,…, n$). As the system changes with time, the representative points trace out a curve in phase space, which is known as a trajectory. The notion of phase space can be generalized to geographical field. Using two time series, say, a time series of urban area and that of urban population, to make a Cartesian coordinate system, we will have a very simple phase space pattern.

3. Order space. This kind of space is defined by referring to the definition of phase space. However, it is not characterized by time series data, but by hierarchical data or cross-sectional data, including the geographical data based on rank and order (level). In theory, the function based on *rank* and that based on *order* can be transformed into each other (Chen and Zhou, 2006). For example, let $k$ be city rank based on population size ($k=1, 2, 3 …$). Suppose $P(k)$ refers to the population of the $k$th city, and $S(k)$ to the urban area of the corresponding city. $P(k)$ and $S(k)$ can support a order space in a Cartesian coordinate system.

The three types of space defined for geographical systems are independent of one another. However, there exists translational symmetry between one space and another one. For instance, if we have a map or an RSI of a city, we can examine the city in real space, turning the sample data into Cartesian coordinates if necessary. If we have more than one time series based on different measures (say, area, population) of the city, we can examine the city in phase space. Further, if we have some kind of data of a system of cities, and the city number is great enough, we can examine the cities in order space in given time. It should be pointed out that the concept of phase space defined in this paper is somewhat different from that in physics. In fact, the concept of phase space in physics encompasses both the concept of phase space and that of order space defined here. The reason why geographical space is classified into three types is that we have three sets of fractal dimension corresponding to space, time, and class respectively (Table 1).

Table 1 The three type of geographical spaces—RPO framework

| Space type | Attribute | Allometry | Phenomenon (example) | Relation |
|---|---|---|---|---|
| Real space (R-space) | Space | Dilation allometry | Form, network, pattern, spatial distribution, etc. | $y(r) \propto x(r)^\alpha$ |
| Phase space (P-space) | Time | Longitudinal allometry | Process, evolution, etc. | $y(t) \propto x(t)^\alpha$ |
| Order space (O-space) | Class (Hierarchy) | Cross-sectional allometry | Size distribution, hierarchical distribution, etc. | $y(k) \propto x(k)^\alpha$ |

**Note**: In the formulae, $r$ refers to radius, $t$ to time, and $k$ to rank or order.

### 2.3. Fractal dimensions of the different spaces

Now, we can take systems of cities as an example to illuminate the dimensions of different geographical spaces. It has been demonstrated that a hierarchy of cities can be characterized with a set of exponential functions based on the top-down order $m$ (Chen and Zhou, 2006), namely

$$f_m = f_1 r_f^{m-1}, \tag{9}$$

$$P_m = P_1 r_p^{1-m}, \tag{10}$$

$$S_m = S_1 r_s^{1-m}, \tag{11}$$

where $m=1, 2, \ldots, M$ is ordinal number ($M$ is the class number), $f_m$ represents the number of cities at a given order/level, $P_m$ denotes the mean population size of order $m$, and $S_m$ is the mean area of the $f_m$ cities, corresponding to the mean population $P_m$. The ratio parameters can be expressed as below: $r_f = f_{m+1}/f_m$, $r_p = P_m/P_{m+1}$, $r_s = S_m/S_{m+1}$. As for the coefficients, $f_1$ refers to the number of the primary city ($f_1 = 1$), $P_1$ to the population of the largest city, and $S_1$ to the urban area of the top-class city.

A set of power laws can be derived from equations (9) to (11). Starting from equations (9) and (10), a scaling relation can be obtained as

$$f_m = \mu P_m^{-D}, \qquad (12)$$

in which $\mu = f_1 P_1^D$, $D = \ln r_f / \ln r_p$. Equation (12) can be termed as "size-frequency relation", and $D$ is just the 'fractal dimension' of urban hierarchies. Similarly, combining equations (9) and (11) yields

$$f_m = \kappa S_m^{-v}, \qquad (13)$$

where the parameters may be written as $\kappa = f_1 S_1^v$, $v = \ln r_f / \ln r_s$. Equation (13) can be named "area-frequency relation". Further we can derive the well-known allometric scaling relation between urban area and population from equations (10) and (11) such as

$$S_m = \eta P_m^b, \qquad (14)$$

where $\eta = S_1 P_1^{-b}$, $b = \ln r_s / \ln r_p$. Equation (14) can be called "area-population relation", and it is equivalent to the allometric growth law of the urban area and population (Batty and Longley, 1994; Chen, 2010a; Lee, 1989; Lo, 2002; Woldenberg, 1973).

The scaling laws of urban hierarchies are involved with three basic measures: number, $f$, size, $P$, and area, $S$. Number is associated with network of cites, size with urban population, and area with urban form. Suppose that the dimensions corresponding to number measure, size measure, and area measure are $D_f$, $D_p$, and $D_s$, respectively. According to the relation of geometrical measures, we have

$$f_m \propto P_m^{-D_f^{(o)}/D_p^{(o)}}, \qquad (15)$$

where $D_f^{(o)}$ is the fractal dimension based on the measure $f$, and $D_p^{(o)}$ the dimension on the measure $P$, both of them are defined in order space (Table 2). Comparing equation (15) with equation (12) gives

$$D = \frac{\ln r_f}{\ln r_p} = \frac{D_f^{(O)}}{D_p^{(O)}} \to \frac{D_f^{(R)}}{D_p^{(R)}}. \tag{16}$$

where $D_f^{(R)}$ is the fractal dimension corresponding to the measure $f$, and $D_p^{(R)}$ the fractal dimension corresponding to the measure $P$ defined in the real space. The notation of arrow "→" means "approach to". Both $D_f^{(R)}$ and $D_p^{(R)}$ can be estimated with the radius-number scaling relation, or box-counting method or grid method in real space (Batty and Longley, 1994; Benguigui *et al*, 2000; Feng and Chen, 2010a; Frankhauser, 1998). Similarly we get

$$S_m \propto P_m^{D_s^{(O)}/D_p^{(O)}} \tag{17}$$

and

$$f_m \propto S_m^{-D_f^{(O)}/D_s^{(O)}}, \tag{18}$$

where $D_s^{(o)}$ is the fractal dimension corresponding to $S$. Then we find the relationships between dimensions and the scaling exponent $v$ as well as $b$, which can be formulated as

$$v = \frac{\ln r_f}{\ln r_s} = \frac{D_f^{(O)}}{D_s^{(O)}} \to \frac{D_f^{(R)}}{D_s^{(R)}}, \tag{19}$$

and

$$b = \frac{\ln r_s}{\ln r_p} = \frac{D_s^{(O)}}{D_p^{(O)}} \to \frac{D_s^{(R)}}{D_p^{(R)}}, \tag{20}$$

where $D_s^{(R)}$ is the dimension corresponding to $S$ in the real space.

According to ergodicity theory (Harvey, 1971; Walters, 2000) and some empirical evidences (Chen and Lin, 2009), equations (9) to (11) can be generalized to time domain in terms of symmetry analysis (Chen, 2009). Then, based on a bottom-up order, we have

$$f_T = f_1 r_f^{1-T}, \tag{21}$$

$$P_T = P_1 r_p^{T-1}, \tag{22}$$

$$S_T = S_1 r_s^{T-1}, \tag{23}$$

where $T$ denotes the level of 'urban age' indicative of temporal dimension of urban evolution (Vining, 1977), $f_T$ refers to the number of cities of age $T$ ($T$=1, 2, .., ), and $P_T$ and $S_T$ to the mean population size and mean area of the $f_T$ cities, respectively. The parameters can be expressed as follows: $r_f = f_T/f_{T+1}$, $r_p = P_{T+1}/P_T$, $r_s = S_{T+1}/S_T$. As for the coefficients, $f_1$ refers to the number of the smallest cities, $P_1$ and $S_1$ to the average population and area of the $f_1$ cities.

Three power laws can be derived from equations (21) to (23). Corresponding to equation (12), the first one is

$$f_T = \mu P_T^{-D}, \quad (24)$$

in which $\mu=f_1 P_1^D$, $D=\ln r_f/\ln r_p$. Corresponding to equation (13), the second one is

$$f_T = \kappa S_T^{-v}. \quad (25)$$

where $\kappa=f_1 S_1^v$, $v=\ln r_f/\ln r_s$. Corresponding to equation (14), the third one is

$$S_T = \eta P_T^b, \quad (26)$$

where $\eta=S_1 P_1^{-b}$, $b=\ln r_s/\ln r_p$. This is also the longitudinal allometric relation between urban area and population. Equation (14) can be associated with the cross-sectional allometry, while equation (26) represents the longitudinal allometry. Both equation (14) and equation (26) are concrete forms of equation (3). By equations (24) to (26), we can derive three dimensions of phase space: $D_f^{(P)}$ refers to the dimension corresponding to number $f$, $D_p^{(P)}$ to the dimension corresponding to size $P$, and $D_s^{(P)}$ to the dimension corresponding to area $S$. All of these dimensions are tabulated here for the purpose of comparison (Table 2).

Table 2 Various dimensions of different types of geographical spaces

| Measure | Geographical systems | | Dimensions of different spaces | | | |
|---|---|---|---|---|---|---|
| | Cities | Rivers | R-space | P-space | O-space | Euclidean |
| Number | Number ($f$) | Number ($N$) | $D_f^{(R)}$ | $D_f^{(P)}$ | $D_f^{(O)}$ | $d_f$ |
| Size | Population ($P$) | Length ($L$) | $D_p^{(R)}$ | $D_p^{(P)}$ | $D_p^{(O)}$ | $d_p$ |
| Area | Area ($S$) | Area ($A$) | $D_s^{(R)}$ | $D_s^{(P)}$ | $D_s^{(O)}$ | $d_s$ |

The real space is easy to understand, and the phase space is familiar to some geographers. However, the order space is first defined in this paper. Although order space bears an analogy with phase space, and in practice, order space is always regarded as phase space, it is necessary for us to make the concept clear in geographical context. For illustrating the three types of geographical spaces and the corresponding dimensions, several examples are presented in next section.

## 3. Examples

### 3.1. An example of real space

The first example is about the real space of Beijing city, the national capital of China, the fractal dimension of which can be directly computed by area/number-radius scaling relation. We can take the urban population and land use area as two measures to show how to estimate fractal dimension as well as the allometric scaling exponent. The fifth census dataset of China in 2000 and the land use dataset of Beijing in 2005 are available. The urban growth core of Beijing, Qianmen, is taken as the center, and a set of concentric circles are drawn at regular intervals. The width of an interval represents 0.5 kilometers (km) on the earth's surface. The land use area between two circles can be measured with the number of pixels on the digital map, and it is not difficult to calculate the area with the aid of ArcGIS software. In this way, the land use density can be easily determined. The population within a ring is hard to evaluate because the census is taken in units of *jie-dao* (sub-district) and each ring runs through different *jie-dao*s. This problem is resolved by estimating the weighted average density of the population within a ring (Chen, 2010b).

Using the datasets of urban land use area and population, we can estimate the allometric scaling exponent of urban form. For real space, the allometric scaling relation is a typical geometric measure relation. According to the scaling pattern, the geographical space of Beijing city can be divided into two scaling ranges, which remind us of bi-fractals of urban land-use structure (White and Engelen, 1993; White and Engelen, 1994). The bi-fractal pattern is often related with self-affine growth of a city. In the first scaling range (0~7km), the allometric scaling relation is

$$S_1(r) = 1.263 P_1(r)^{1.121},$$

where $r$ refers to the distance from the center of the city ($r$=0), $P(r)$ to the population within a radius of $r$ km of the city center, $S(r)$ to the built-up area of urban land use. The goodness of fit is about $R^2$=0.999. In the second scaling range (7.5-48km), the allometric scaling relation is

$$S_2(r) = 45.825 P_2(r)^{0.417}.$$

The goodness of fit is around $R^2$=0.976 (Figure 1).

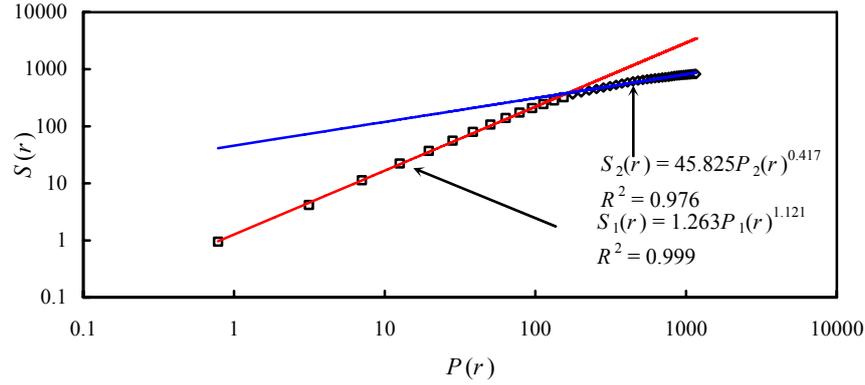

**Figure 1 The log-log plot of the scaling relation between urban area (in 2005) and population (in 2000) of Beijing** (The unit of land use area is km, and population, 10 thousands)

The first model reflects the spatial feature of urban core (0~7km), and the allometric scaling exponent is about $b$=1.121. This suggests that urban land use becomes denser the more distant from the city center it is. This is not normal. The second model implies the human geographical structure of urban periphery and suburban area, and the allometric scaling exponent is about $b$=0.417. This implies that urban land use becomes less dense the more distant from the city center you go. But the $b$ value is too low. The line of demarcation between the two scaling ranges appears at about 7 kilometer. In fact, for the real space, the fractal dimension of Beijing's urban population distribution and land use can be directly estimated by using the area/number-radius scaling. This kind of dimension is called *radial dimension* (Frankhauser and Sadler, 1991; White and Engelen, 1993). For the urban land use in 2005, the fractal model is

$$S(r) = 3.534 r^{1.904}.$$

The goodness of fit is about $R^2$=0.997, and the radial dimension is $D_s$≈1.904 (Figure 2(a)). For the spatial distribution of urban population in 2005, the fractal model is

$$P(r) = 8.609 r^{1.616}.$$

The goodness of fit is around $R^2$=0.935, and the radial dimension is $D_p$≈1.616 (Figure 2(b)). The allometric scaling exponent of Beijing from 0 to 48 km can be approximately estimated as $b$≈1.616/1.904≈0.849.

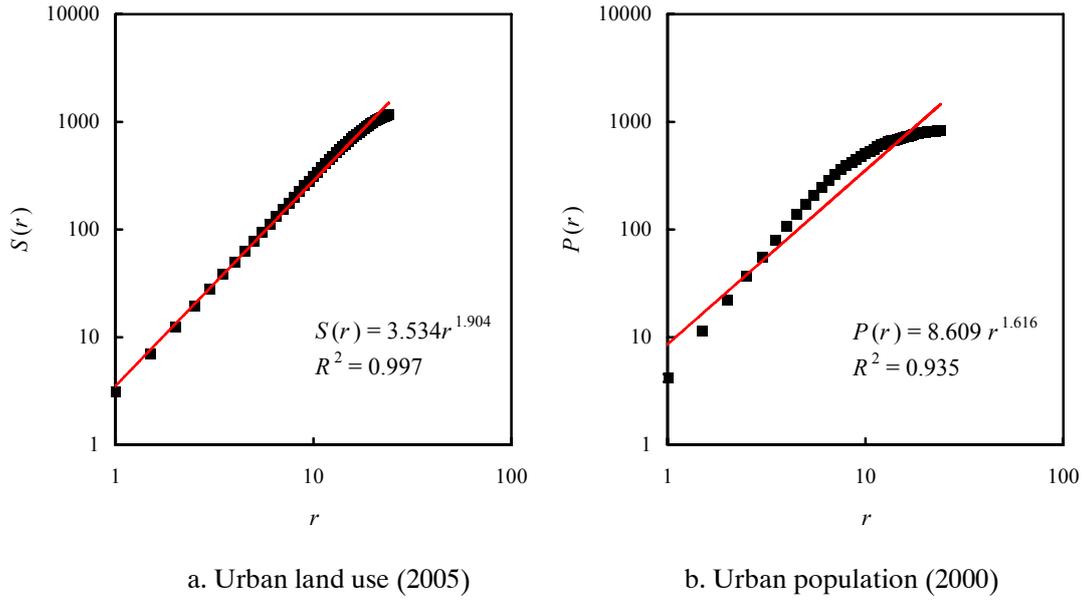

a. Urban land use (2005)  b. Urban population (2000)

**Figure 2 The log-log plots of the urban land use area and population of Beijing in 2000 and 2005**

This case analysis of Beijing is deficient in preciseness because of two shortcomings. First, the years of sampling are not consistent with each other. The population dataset is in 2000, whereas the land use dataset is in 2005. Consider the time lag of the correlation between urban population and land use, the inconsistent sampling is acceptable, or tolerable at least. After all, the concentration and diffusion of urban population is previous to land use. Second, as stated above, the population distribution cannot be modeled by the self-similar fractal based on the number-radius scaling. Obviously, the data of urban population cannot be well fitted to the number-radius scaling relation. Maybe the population distribution follow the exponential law instead of the power law, maybe the spatial distribution of population is of self-affinity rather than self-similarity. Just because of this, the allometric scaling relation of urban area and population broke up into two scaling ranges (Figure 1). Despite these defects, it is sufficient to show the allometric scaling relation and fractal dimensions of the real space.

Actually, fractal dimension values depend on the measurement method. If we use the box-counting method to research city fractals, the effect is always satisfying; however, if we employ the area/number-radius scaling to estimate fractal dimension of a city, the result is not often convincing. The fractal dimension of urban form based on the box-counting method comes between 1 and 2. As for Beijing, the box dimension of the area within the frame of viewfinding (study area) is about $D_f^{(R)}=1.90$ in 1999 and $D_f^{(R)}=1.93$ in 2006 (Figure 3). However, the dimension values based on the area-radius scaling can be greater than 2 or less than 1 (White and Engelen, 1993; White, *et al*, 1997). According to Frankhauser (1994), the fractal dimension of Beijing's city form is about 1.93 in 1981 (see Batty and Longley, 1994, p242). Today, the radial dimension value is greater than $d=2$ (Jiang and Zhou, 2006). If we examine the population in the region within a radius of 7 km of the city center, the model is $P(r)=4.555r^{2.242}$, the goodness of fit is about $R^2=0.999$. The topic of this paper is not fractal study of urban form, but classification of geographical spaces and dimensions. The city of Beijing is only taken here as an example to show what is the fractal dimension of the real space.

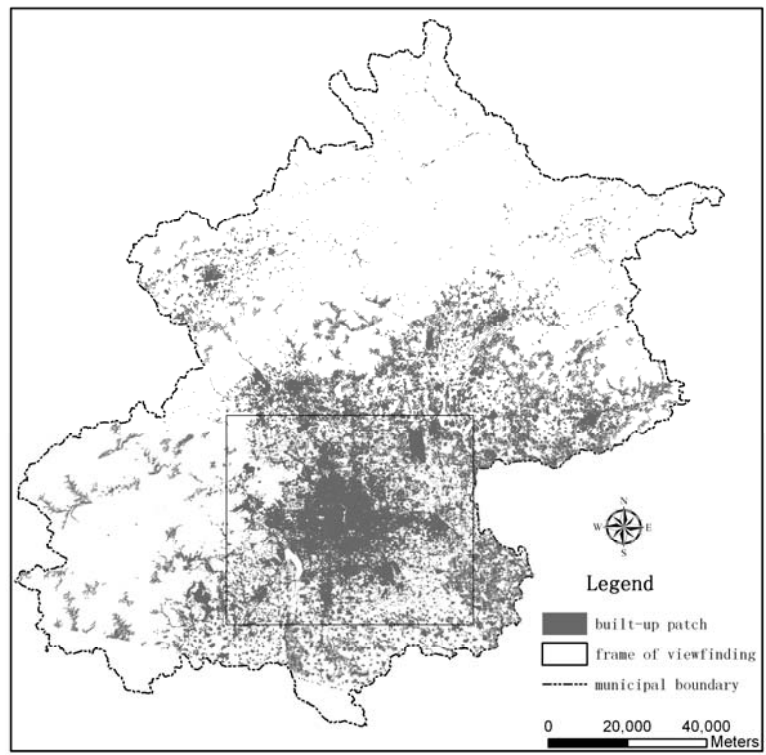

**Figure 3 The urban fractal landscape of Beijing in 2006**

### 3.2. An example of phase space

The second example is on the phase space of Zhengzhou city, the capital of China's Henan Province. It is hard to calculate its fractal dimension directly. Fortunately, the ratio of one dimension to another can be estimated through the allometric scaling relation (Chen and Jiang, 2009). In practice, the fractal dimension ratio is more important than fractal dimension values themselves (Chen and Lin, 2009). The variables are urban area and population, and the two sample paths of time series range from 1984 to 2004. In other words, we have two observational datasets: one is for urban area, and the other, for population size. The linear relation between urban area and population is clear in the double logarithmic plot (Figure 4). A least squares computation give the following result

$$S(t) = 0.331 P(t)^{1.163},$$

where $t$ refers to time (or year). The goodness of fit is $R^2=0.969$. This implies that the ratio of fractal dimension of urban form to that of urban population in phase space is as follows

$$b = \frac{D_s^{(P)}}{D_p^{(P)}} = 1.163.$$

The fact that the dimension of urban form is greater than that of urban population suggests that urban land use grew faster than urban population (Chen, 2010a; Lee, 1989).

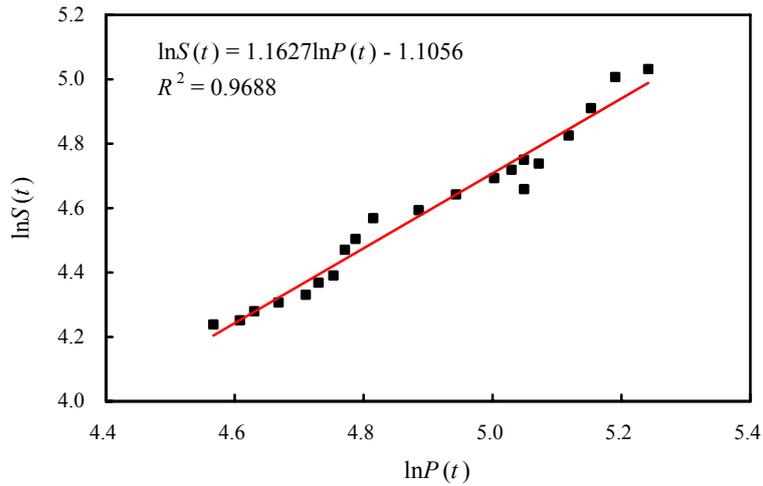

Figure 4 The log-log plot of longitudinal allomtric relation between urban area and population of Zhengzhou, 1984-2004

Generally speaking, it is almost impossible for us to evaluate the capacity dimension ($D_0$) and information dimension ($D_1$) of urban area and population in phase space, but it is very easy for us to estimate the correlation dimension ($D_2$) by reconstructing phase space with time series (Figure 5). In a multifractals dimension spectrum ($D_q$, where $q$ denotes the moment order, and $-\infty<q<\infty$), the correlation dimension is less than the capacity dimension and information dimension (Feder, 1988). However, the ratio of one correlation dimension to another correlation dimension can be probably brought into comparison with the ratio of one capacity dimension to another capacity dimension. If the dimension of embedding space $d_e=2$ as given, then the correlation dimension value of urban area in the reconstructed phase space of time series from 1984 to 2004 is estimated as $D_{s2}^{(P)} \approx 0.843$, the goodness of fit is about $R^2=0.996$ (Figure 3a). Accordingly, the correlation dimension of urban population is about $D_{p2}^{(P)} \approx 0.773$, the goodness of fit is around $R^2=0.995$ (Figure 3b). The ratio of correlation dimension of urban area to that of urban population is

$$b_2 = \frac{D_{s2}^{(P)}}{D_{p2}^{(P)}} = \frac{0.843}{0.773} = 1.091,$$

where the subscript 2 implies correlation dimension, namely, the order of moment is $q=2$. This value ($b_2=1.091$) is consistent with the result given by the longitudinal allometric relation ($b=1.163$).

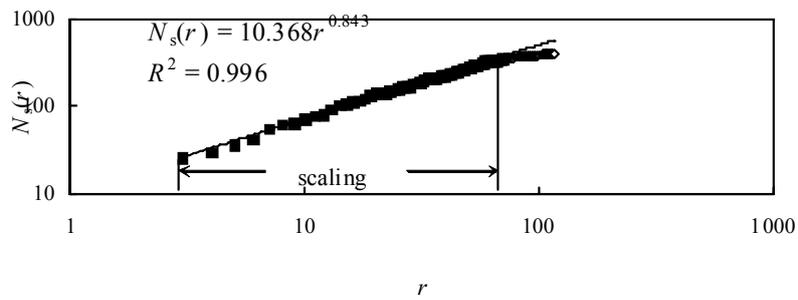

a. Urban area correlation ($d_e=2$)

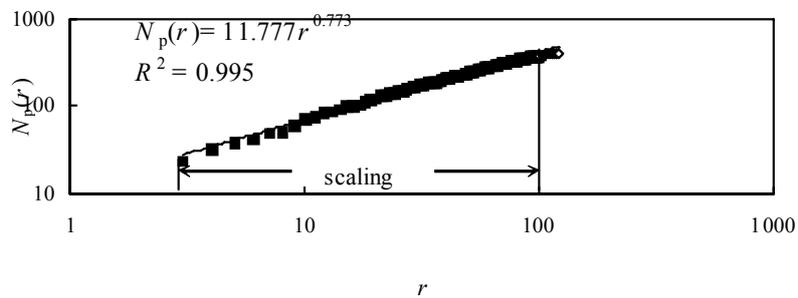

b. Urban population correlation ($d_e=2$)

**Figure 5 The scaling relationships between urban area and population of Zhengzhou in the reconstructed phase space (1984-2004)**

### 3.3. An example of order space

The third example is on the order space of China's urban system consisting of 664 cities in 2000, the dimension of which cannot be directly evaluated, either. The data can be processed through two approaches, which are equivalent to each other in theory (Chen and Zhou, 2006). The first is based on rank of size distribution, and the other on order of self-similar hierarchy. Based on the rank, the mathematical model is

$$S_k = 1.927 P_k^{0.812},$$

where $k$ denotes city rank ($k=1,2,\cdots,664$), the goodness of fit is $R^2=0.776$ (Figure 6(a)). This implies that the ratio of fractal dimension of urban form to that of urban population in order space is as follows

$$b = \frac{D_s^{(O)}}{D_p^{(O)}} = 0.812.$$

By the method of hierarchical scaling and rescaling, the city size distribution can be converted into a self-similar hierarchy (Chen, 2009; Chen, 2010a). Based on the hierarchical order, the model is

$$S_m = 1.881 P_m^{0.843},$$

where $m$ indicates urban order in the hierarchy ($m=1,2,\cdots,9$), the goodness of fit is $R^2=0.994$ (Figure 4(b)). This suggests a fractal dimension ratio such as

$$b = \frac{D_s^{(O)}}{D_p^{(O)}} = 0.843.$$

Apparently, the two results, 0.812 and 0.843, are close to one another empirically.

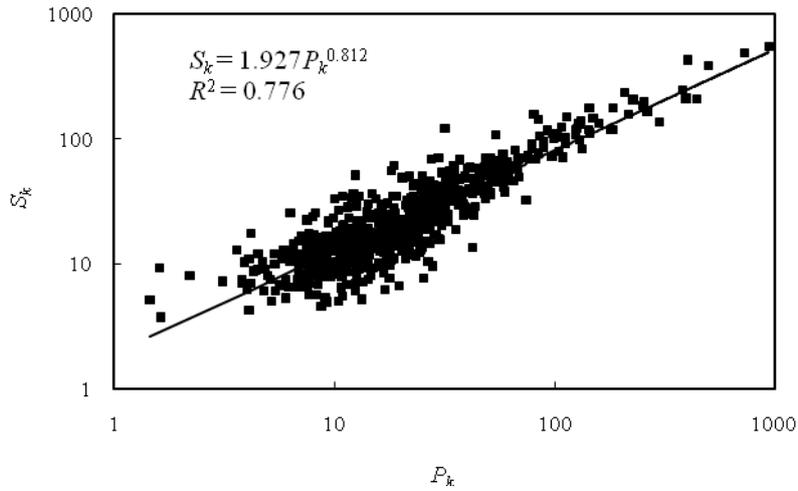

a. Based on rank

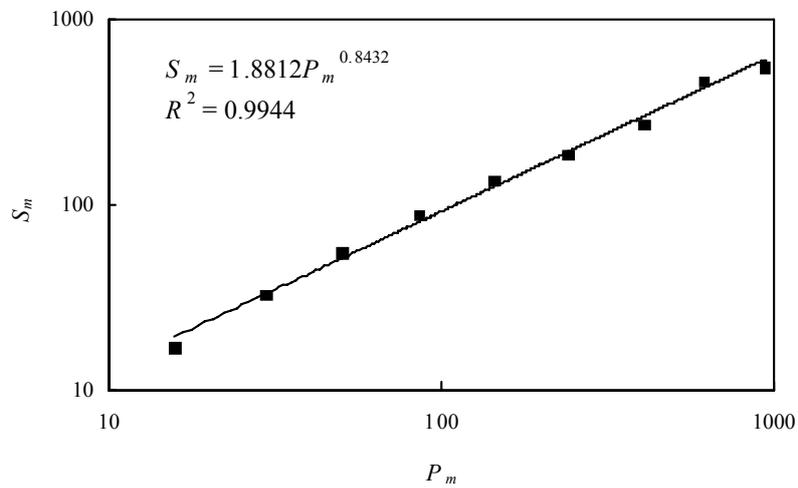

b. Based on order

**Figure 6 The cross-sectional allometric relation between urban area and population of China's cities in 2000 (Data sources:** (1) Population of city regions: 2000 census of China; (2) Build-up area: Ministry of Construction, P. R. China. 2000 *Statistic Annals of China's Urban Construction*. BJ: Chinese Architecture Industry Publishing Housing, 2001.)

## 4. Applications

### 4.1. Explanations of river and city models

The theory of RPO spaces and dimensions proposed in this paper can be employed to explain many traditional models and related parameters in both physical and human geography. In fractal studies of both river networks and urban hierarchies, there exist some difficulties in interpreting the meaning of fractal parameters as well as the differences between calculated results and theoretically expected values. The crux of the matter lies in the fact that the fractal dimensions based on different kinds of spaces are easily confused with each other. Actually, the generalized fractal dimension of order space or phase space (say, similarity dimension) is always treated as the dimension of real space (say, box dimension). Once the geographical spaces and dimensions are put in order, many problems can be readily resolved. Two examples are presented here to illuminate these questions.

Hierarchies of cities have been demonstrated to share the same scaling laws with networks of rivers (Chen and Zhou, 2006; Chen, 2009). Suppose that the rivers in a system are divided into $M$ levels in a bottom-up order. The structure of system of rivers can be formulated as three exponential laws as follows

$$N_m = N_1 r_b^{1-m}, \tag{27}$$

$$L_m = L_1 r_l^{m-1}, \tag{28}$$

$$A_m = A_1 r_a^{m-1}, \tag{29}$$

where $m$ is the increasing order of the hierarchy of rivers ($m=1, 2, \cdots, M$), $N_m$ is the number of river branches of order $m$, $L_m$ is the mean length of the branches of order $m$, and $A_m$ is the mean catchment area of the rivers of order $m$ corresponding to the mean length $L_m$. As for the parameters, $N_1$ is the number of the first-order branches, $r_b=N_m/N_{m+1}$ is termed *bifurcation ratio*; $L_1$ is the mean length of first-order rivers, $r_l=L_{m+1}/L_m$ is the *length ratio*; $A_1$ is the mean catchment area of the first-order rivers, and $r_a=A_{m+1}/A_m$ is the *area ratio*. The formula of bifurcation ratio, length ratio, and area ratio are actually three translational scaling laws taking on the form of geometric series (Williams, 1997). Equations (27) to (29) are just the law of river composition originated by Horton (1945) and consolidated by Strahler (1952) and then developed by Schumm (1956).

It has been shown that river systems have fractal property (LaBarbera and Rosso, 1989; Rodriguez-Iturbe and Rinaldo, 2001; Tarboton, *et al*, 1988; Turcotte, 1997). Turning the bottom-up order into the top-down order based on the special symmetry of exponential function (Chen, 2009), a number of power functions can be derived from equations (27) to equation (29).

Combining equation (27) with equation (28) yields

$$N_m = kL_m^{-D}, \quad (30)$$

where $k=N_1L_1^D$, $D=\ln r_b/\ln r_l$, and $D$ is regarded as the fractal dimension of river systems. Similarly, combining equation (28) with equation (29) gives an allometric scaling relation

$$L_m = cA_m^h, \quad (31)$$

in which $c=L_1A_1^{-b}$, $h=\ln r_l/\ln r_a$, This is actually the generalized Hack's law (Hack, 1957), and the scaling exponent is considered to be relative to fractal dimension (Mandelbrot, 1983; Feder, 1988). Combining equation (27) with equation (29) yields

$$N_m = uA_m^{-\sigma}, \quad (32)$$

where $u=N_1A_1^\sigma$, $\sigma=\ln r_b/\ln r_a$. Obviously, we have

$$\sigma = D \cdot h = \frac{\ln r_b}{\ln r_a}. \quad (33)$$

In light of the Korcak's law (Mandelbrot, 1983), $\sigma$ approaches to 1, then $r_a \approx r_b$, and thus we have $h=\ln r_l/\ln r_b$ (Feder, 1988). However, this is not often the case in the real world (Table 2).

The fractal models of rivers, equations (30) to (32), gave rise to two difficult problems in both theory and practice. First, we should clarify the problem related to equation (30). The parameter in equation (30), $D$, used to be taken as the fractal dimension equivalent or even identical to the box dimension indicating real space measurement. Thus the value of $D$ should range between 1 (the topological dimension of a river) and 2 (the embedding space dimension of rivers). However, the scaling exponent values given by equation (30) are not always consistent with the fractal dimension values from the box-counting method. What is more, the $D$ value is sometimes beyond the limits 1 and 2 (Table 3). Geographers are always in a puzzle about this kind of phenomena.

The abovementioned difficult problem can be readily solved by using the theory of this paper. In fact, the parameter $D=\ln r_b/\ln r_l$ is not really a fractal dimension defined in real space, but a generalized fractal dimension ratio defined in order space. In terms of geometrical measure relation, we have

$$N_m \propto L_m^{-D} \propto L_m^{-D_n^{(O)}/D_l^{(O)}}. \quad (34)$$

That is

$$D = \frac{D_b^{(O)}}{D_l^{(O)}} = \frac{\ln r_b}{\ln r_l} \to \frac{D_b^{(R)}}{D_l^{(R)}}, \quad (35)$$

where $D_b^{(O)}$ refers to the dimension of a network of rivers, and $D_l^{(O)}$ to the dimension of watercourses in the order space. The former corresponds to the structural fractal dimension of river networks in real space, $D_b^{(R)}$, which can be evaluated by the box-counting method; the latter corresponds to textural fractal dimension of riverways in real space, $D_l^{(R)}$, which can be estimated by the Richardson (1961)'s divider method. Generally speaking, $2>D_b^{(R)}>1.25$, $1.75>D_l^{(R)}>1$; therefore we have $2>D>1$ frequently. However, as a fractal dimension ratio defined in the order space, it is not surprising for $D$ to be less than 1 or greater than 2. In short, $D=\ln r_b/\ln r_l$ is the fractal dimension ratio of hierarchy of rivers in the order space rather than that of network of rivers in the real space.

**Table 3 The values of fractal parameter ($D$), scaling exponent ($h$), and some related parameters of river systems in Jilin Province, China**

| River | $D=\ln r_b/\ln r_l$ | $R^2$ | $h=\ln r_l/\ln r_a$ | $R^2$ | $2h\approx D_l$ | $1/h$ | $\sigma=Dh$ | $2\sigma\approx D_a$ |
|---|---|---|---|---|---|---|---|---|
| The first Songhua river | 1.619 | 0.978 | 0.583 | 0.988 | 1.166 | 1.715 | 0.944 | 1.888 |
| The second Songhua river | 1.606 | 0.899 | 0.563 | 0.988 | 1.126 | 1.776 | 0.904 | 1.808 |
| Huifa river | 1.630 | 0.972 | 0.543 | 0.970 | 1.086 | 1.842 | 0.885 | 1.770 |
| Yinma river | 1.343 | 0.962 | 0.610 | 0.941 | 1.220 | 1.639 | 0.819 | 1.638 |
| Taoer river | *0.759* | 0.878 | 0.636 | 0.988 | 1.272 | 1.572 | 0.483 | *0.965* |
| Eastern Liao river | 1.274 | 0.863 | 0.627 | 0.998 | 1.254 | 1.595 | 0.799 | 1.598 |
| Hun river | 1.425 | 0.970 | 0.561 | 0.988 | 1.122 | 1.783 | 0.799 | 1.599 |
| Peony river | 1.549 | 0.978 | 0.536 | 0.992 | 1.072 | 1.866 | 0.830 | 1.661 |
| Gaya river | 1.736 | 0.799 | 0.509 | 0.990 | 1.018 | 1.965 | 0.884 | 1.767 |

**Source of the original data**: Water Conservancy Office of Jin Province, China (1988). *Feature Values of Rivers' Drainage Basins in Jilin Province*. In the table, $R^2$ represents the goodness of fit. The data were processed by Dr. Baolin Li of Chinese Academy of Sciences.

Next, the questions from Hack's law should be made clear (Hack, 1957). Generally, equation (31) is re-expressed as a geometrical measure relation such as

$$L_m \propto A_m^{D_l/2}, \quad (36)$$

where $D_l = 2h$ denotes the fractal dimension of watercourses. $D_l$ is originally expressed as (Feder, 1988)

$$D_l = 2h = \frac{2\ln r_l}{\ln r_a} = \frac{2\ln r_l}{\ln r_b}, \qquad (37)$$

which is revised by Tarboton *et al* (1988) as

$$\frac{D_l}{D_a} = h = \frac{\ln r_l}{\ln r_a} = \frac{\ln r_l}{\ln r_b}. \qquad (38)$$

where $D_a$ indicates the fractal dimension of drainage area form. This is apparently a progress in theory, but the problem has not been finally resolved. According to the theory of this paper, we have

$$h = \frac{D_l^{(O)}}{D_a^{(O)}} = \frac{\ln r_l}{\ln r_a} \to \frac{D_l^{(R)}}{D_a^{(R)}}, \qquad (39)$$

$$\sigma = \frac{D_b^{(O)}}{D_a^{(O)}} = \frac{\ln r_b}{\ln r_a} \to \frac{D_b^{(R)}}{D_a^{(R)}}, \qquad (40)$$

where $D_a^{(O)}$ refers to the dimension of the drainage area in the order space corresponding to the fractal dimension of the drainage basin form in the real space, $D_a^{(R)}$, which is expected to equal 2. Thus, we have $D_l^{(O)}=2h$, $D_b^{(O)}=2\sigma$. However, this is not often the case in the real world. For the real rivers, $2\sigma$ values are always greater than $D_b^{(O)}$ values (Table 3). These phenomena cannot be interpreted by the real space theory, but it can be explained with the order space theory.

In urban studies, we have met with the similar problems. The scaling exponent in equation (14), *b*, is hard to understand according to the traditional theory (Lee, 1989). In light of Euclidean geometry, *b* values should be 2/3 or 1. However, *b* values often vary between 2/3 and 1, and the mean of *b* values approaches to 0.85. In fact, *b* is not a fractal parameter defined in real space, but a fractal dimension ratio defined in phase space (for the *longitudinal allometry*) or order space (for the *cross-sectional allometry*). Under the ideal conditions, we have

$$D_p^{(O)} \to D_p^{(R)}, D_s^{(O)} \to D_s^{(R)}.$$

According to Batty and Longley (1994), the mean of $D_s^{(R)}$ approaches to 1.7. If population distribution is self-similar, then $D_p^{(R)}$ is expected to near 1.7, thus, according to equation (20), we have $b \approx 1$; if population distribution is not self-similar, the $D_p^{(R)}$ is expected to equal 2, and consequently we have $b=1.7/2=0.85$ (Chen, 2010a). All of these are discussed from the angle of view of statistical average.

## 4.2. Ergodicity and locality

According to the ergodicity hypothesis (Walters, 2000), if a system is ergodic, we have a formula as "time average=space average=phase average". Ergodicity is a contrast to localization (El Naschie, 2000; Liu and Chen, 2007). For a system of non-localization, in theory, we have

$$D^{(R)} = D^{(P)} = D^{(O)}. \tag{41}$$

This relation is true only when the geographical system is ergodic, and under ideal conditions. A real geographical system always has two major, sometimes contradictory, specialities indicative of two different trends: one is ergodicity, the other is locality. Because of locality, the above equality always breaks down and is replaced by a inequality such as

$$D^{(R)} \ne D^{(P)} \ne D^{(O)}. \tag{42}$$

However, ergodicity always appears at the macro level, while locality always appears at the micro level. If a sample is so large that it approximates to the population, or if we take an average from a very large sample, the result is expected to be

$$D^{(R)} \approx D^{(P)} \approx D^{(O)}. \tag{43}$$

That is to say, statistical averages always show ergodicity and screen locality. So, further test of the theory proposed in this paper should be made by using the idea of statistical average.

In practice, fractal dimension values are often difficult to be evaluated, but the ratios of one fractal dimension to another fractal dimension are easily estimated by means of allometric growth or geometric measure relations (Chen, 2010a). On the other hand, in geographical analysis, the relative numerical values of fractal dimension such as fractal dimension ratios are more significant than the absolute numerical values of fractal dimension itself. In this sense, this paper proposes a theoretical basis for geographical research by using fractal dimension ratios based on the macro level. Especially, as a speculation, equations (41) can be used as a criterion of structural optimization of geographical systems.

## 5. Conclusions

Geographers are very familiar with the notion of the real space, and the concept of phase space is familiar to theoretical geographers. However, the idea of order space used to be confused with phase space or even real space. In essence, the dimension of order space differs from that of real space or phase space. A different kind of dimension indicates a different type of space. One of the aims of this paper is to remove confusion over order and real space in geography. Division and classification of geographical space are helpful for us to make theoretical analysis of geographical systems by means of scaling relations.

The main points of this paper can be summarized as follows. According to the sorts of dimensions, geographical space falls into three types, namely, real space (R-space), phase space (P-space), and order space (O-space), which constitutes a RPO framework of geographical space theory. R-space can be represented by maps, images, etc. The dimension values of R-space can be directly computed by box-counting method, area/number-radius scaling, or Richardson's method. P-space can be reflected with time series. The dimension of P-space is hard to be directly evaluated, but the correlation dimension of time series can be calculated by reconstructing phase space. Especially, the ratio of dimensions can be estimated through the longitudinal allometric relations. O-space can be reflected by the data based on rank-size distributions or hierarchical structure. The dimension of O-space is also difficult to be directly evaluated, but the ratio of dimensions can be easily estimated through cross-sectional allometric analysis.

The significance of developing the theory of generalized geographical spaces and dimensions is as follows. In theory, the idea of three-type spaces can be employed to explain the parameters of many scaling laws on geographical systems. In practice, the ratios of fractal dimensions are usually more important than fractal dimensions themselves. The translational symmetry of different kinds of space may be a criterion of system optimization. By means of the concept of RPO spaces, different fractal dimension ratios can be estimated. In virtue of different dimensions and ratios of dimensions, we can characterize geographical systems well and make in-depth analysis of geographical spatial information.

**Acknowledgements**: This research was sponsored by the National Natural Science Foundation of China (Grant No. 41171129) and the Natural Science Foundation of Beijing (Grant No. 8093033). The author would like to thank Ms. Jingyi Lin of Royal Institute of Technology (RIT), Sweden, for providing the essential data on the urban land use and population of Beijing and Dr. Baolin Li of Chinese Academy of Sciences (CAS), China, for providing the results of fractal parameters of Jilin's rivers. The picture of Beijing in 2006 came from Jiejing Wang of Beijing University (PKU), China.